\documentclass[aps,prl,reprint,groupedaddress]{revtex4-1}

\usepackage{graphicx}
\usepackage{amsmath}

\usepackage{color}

\begin{document}


\title{Detecting chirality in molecules by linearly polarized laser fields}



\author{%
Andrey Yachmenev}
\affiliation{%
Department of Physics and Astronomy, University College London, London, WC1E 6BT, UK}

\author{%
Sergei N. Yurchenko}
\affiliation{%
Department of Physics and Astronomy, University College London, London, WC1E 6BT, UK}


\date{\today}

\begin{abstract}

A new scheme for enantiomer differentiation of chiral molecules using a pair of linearly polarized intense ultrashort laser pulses with skewed mutual polarization is presented. The technique relies on the fact that the off-diagonal anisotropic contributions to the electric polarizability tensor for two enantiomers have different signs. Exploiting this property, we are able to excite a coherent unidirectional rotation of two enantiomers with a $\pi$ phase difference in the molecular electric dipole moment. The approach is robust and suitable for relatively high temperatures of molecular samples, making it applicable for selective chiral analysis of mixtures, and to chiral molecules with low barriers between enantiomers. As an illustration, we present nanosecond laser-driven dynamics of a tetratomic non-rigid chiral molecule with short-lived chirality. The ultrafast time scale of the proposed technique is well suited to study parity violation in molecular systems in short-lived chiral states.


\end{abstract}


\maketitle

Among the most important and challenging tasks of chemical research is the detection and measurement of the enantiomeric excess and handiness of chiral molecules~\cite{Bowering01,Nahon06,Li08,He11,Hiramatsu12,Lux12,Patterson13,Pitzer13,Janssen14,Patterson14}, as well as chiral purification and discrimination of their racemic mixture~\cite{Bodenhofer97,McKendry98,Rikken00,Zepik02,Kral03}. Molecular chirality is crucial to much of chemistry and also plays an important role in fundamental physics representing systems with broken (mixed) parity states~\cite{08QuStWi.chirality,HiResBook}. A considerable amount of effort is currently being devoted to observing parity violation in chiral molecules and any experimental detection would be a significant breakthrough.


A number of ground breaking experiments for investigating chiral molecules in the gas phase have recently been developed. These include photoelectron circular dichroism using intense laser pulses or synchrotron radiation~\cite{Bowering01,Lux12,Janssen14}, and an approach based on laser ionization-induced Coulomb explosion imaging~\cite{Pitzer13,Herwig13}. 
The most relevant for the present work, however, is the microwave three-wave mixing experiment~\cite{Patterson13,Patterson14,Shubert14}. Here, the resonant microwave fields are used to couple three selected rotational energy levels of a chiral molecule that are connected by transitions via three different dipole projections. Due to the opposite sign of one of the dipole moment projections between two enantiomers, they evolve with intrinsically different total phases and this can be seen in the phase of the emitted radiation. By measuring the absolute amplitude of the emitted signal, the enantiomeric excess and absolute configuration can be quantified~\cite{Grabow13}.

The major advantages of the gas phase resonant three-wave mixing technique are (i) its high sensitivity to even a tiny enantiomeric excess and (ii) applicability to the analysis of mixtures of different conformers and stereoisomers~\cite{Shubert14}. The disadvantage is that the method can only be used on molecules characterized by three non-zero projections of the the permanent electric dipole (in the principal axis system). Even small dipole moment projections will cause slow Rabi oscillations of populations between two states of a three-level system which would demand long, impractical measurements. As a result, short-lived chirality in low-barrier axially chiral molecules such as H$_2$O$_2$ and H$_2$S$_2$ is extremely difficult to detect.

In this work, we propose a novel scheme to circumvent this problem by involving the non-zero off-diagonal anisotropic polarizabilities of the enantiomers. We demonstrate that a pair of intense ultrashort laser pulses with skewed mutual polarization can be used to induce the enantiomer-specific phase shift in the emitted microwave signals, required for the differentiation of enantiomers. As shown in \citep{Fleischer09,Kitano09}, two laser pulses can produce a unidirectional coherent molecular rotation, which is demonstrated here for two enantiomers. This generates a phase shift in the time evolution of their permanent electric dipole moment due to the opposite signs of the corresponding off-diagonal anisotropic polarizabilities. The method requires at least one non-zero projection of the permanent electric dipole moment and anisotropic electric polarizability. We also show that by adjusting the time delay between the two pulses, the amplitude of the dipole oscillations can either be increased or decreased. Such an effect can be exploited in the chiral analysis of mixtures of different molecular species. Since the excitation wavelengths of the present scheme are not tuned to the specific transitions of the system, it is more robust with respect to increasing temperature. The characteristic time of the scheme is on the picosecond scale, which is also favourable for measurements of short-lived chirality.

Let us first consider two linearly polarized non-resonant laser fields sequentially applied to a chiral molecule satisfying the conditions above.
The beam propagates in the direction of the $Z$ axis, the polarization vector of the first pulse is set along the $X$ axis, and the second pulse, applied with a time delay of $\tau$ ps, is linearly polarized with the angle $\gamma$ to the first pulse.
The interaction potential of an enantiomer with the electric fields  in the absence of electronic resonances can be described in the spherical tensor form  as given by 
\begin{eqnarray}\label{eq1}
V(t) &=& \frac{1}{4}E(t)^2\left(\frac{2}{\sqrt{3}} D_{00}^{(0)*}\alpha_{0}^{(0)}+\sqrt{\frac{2}{3}}D_{0k}^{(2)*}\alpha_{k}^{(2)}\right. \\ \nonumber
&-&\left. [D_{-2k}^{(2)*}e^{2i\gamma}+D_{2k}^{(2)*}e^{-2i\gamma}]\alpha_{k}^{(2)}\right) .
\end{eqnarray}
Here, $\alpha_{k}^{(J)}$ are the spherical tensor components of the molecular electric polarizability in the molecule fixed frame, $D_{mk}^{(J)*}$ is the complex conjugate Wigner rotation matrix, $E(t)$ is a time-dependent external electric field, and implicit summation over the repeated index $k=-2..2$ is assumed.
The external electric field has the form $E(t)=\varepsilon(t;t_0,T)\cos \omega t$, where $\varepsilon(t;t_0,T)$ is the pulse Gaussian time profile with a maximum at $t=t_0$ and a duration (FWHM) $T$, and $\omega$ is the carrier frequency.
The spherical tensor components of the molecular polarizability can be represented in terms of real Cartesian components in the molecule fixed frame by $\alpha_{0}^{(0)}=-\frac{1}{\sqrt{3}}(\alpha_{xx}+\alpha_{yy}+\alpha_{zz})$, $\alpha_{0}^{(2)}=\frac{1}{\sqrt{6}}(2\alpha_{zz}-\alpha_{xx}-\alpha_{yy})$, $\alpha_{\pm 2}^{(2)}=\frac{1}{2}(\alpha_{xx}-\alpha_{yy} \pm 2 i \alpha_{xy})$, and $\alpha_{\pm 1}^{(2)}=\mp \alpha_{xz}- i \alpha_{yz}$.

As one can see, the interaction potentials $V(t)$ of two enantiomers differ by signs of the two off-diagonal Cartesian polarizability elements in the molecule fixed frame.
This implies that an electric field will create excitations with a nonuniform distribution of projections $k$, of total angular momentum onto a molecule fixed axis $z$, such that the populations of $-k$ of one enantiomer will be equal to the populations of $+k$ of another enantiomer, and vice versa.
The `counter' distribution of projections $k$ in the wave packets of two enantiomers will thus produce completely antiphase time-evolution modulations of the expectation value of the laboratory fixed dipole moment operator.

In the absence of an electric field, the isotropy of a molecular system is represented by a uniform distribution of projections $m$, of total angular momentum onto a laboratory fixed axis $Z$.
The matrix elements of $V(t)$ in the field-free basis are non-vanishing for $\Delta m=0,\pm 2$ only (electric polarizability selection rules) and the rotational transition probabilities for $\Delta m=\pm 2$ are the same to each other. The first ultrashort pulse induces a molecular alignment by creating a packet with a non-uniform distribution of excitations characterized by different projections $|m|$ but the same populations of states  with $-m$ and $+m$.
Due to the axial symmetry of excitation, matrix elements of the laboratory fixed dipole moment operator will vanish.
In order to allow the dipole transitions we need to break the axial symmetry. This is done by applying the second pulse, which is linearly polarized at the angle $\gamma$ to the first one. As seen from the second row of Eq.~(\ref{eq1}) for $V(t)$, the maximal asymmetry with respect to $\Delta m=\pm 2$ transition probabilities can be achieved for the angle $\gamma = \pm 45^\circ$, and the axial symmetry is fully preserved for  $\gamma=0$ or $\pm 90^\circ$. The second pulse thus creates a non-uniform distribution of excitations with $+m$ and $-m$, which can lead to a non-zero expectation value of the dipole moment operator.
In turn, the dipole oscillations from two enantiomers produce microwave signals with $\pi$ phase difference and the total amplitudes, in the case of excess of one enantiomer over the another, can be amplified and detected~\cite{Patterson13}.

\begin{figure}
\includegraphics[scale=0.20]{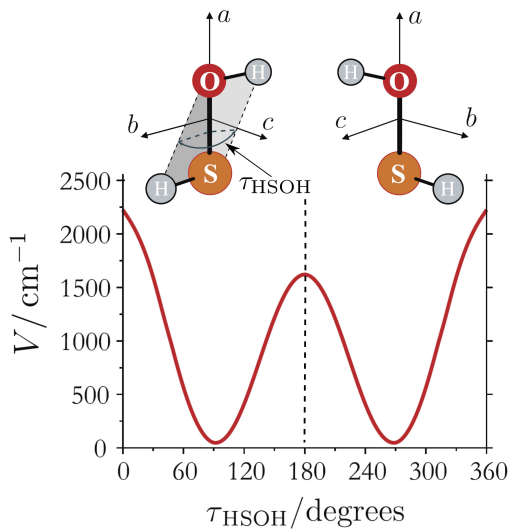}%
\caption{\label{fig:hsoh}
One-dimensional cut through the PES of HSOH calculated along the torsional coordinate $\tau_{\rm HSOH}$ together with the definition of the $abc$ molecule fixed axes system for two enantiomeric molecular configurations.}
\end{figure}

We now turn to numerical simulations and consider the closed-shell asymmetric-top molecule HSOH as a working example.
HSOH has six vibrational degrees of freedom with one large amplitude torsional motion of the H atoms about the S--O bond described by the coordinate $\tau_{\rm HSOH}$.
A calculated {\it ab initio} one-dimensional cut of the ground electronic state potential energy surface (PES)~\cite{HSOH_abinit} along $\tau_{\rm HSOH}$ is shown in Fig.~\ref{fig:hsoh}.
The PES has two minima, corresponding to the two axial enantiomeric forms, separated by two barriers of 1520.9~cm$^{-1}$ and 2163.3~cm$^{-1}$~\cite{HSOH_abinit,03QuWixx.HSOH}.
The calculated tunnelling splitting of the lowest torsional states is $\Delta \tilde{E}=0.00215$~cm$^{-1}$~\cite{HSOH_abinit}, which is in perfect agreement with the experimental value of 0.00214~cm$^{-1}$~\cite{HSOH_exp_split}, and predicts the lifetimes of enantiomers of about 8~ns.
Hence, the molecule can `live' in one of the enantiomeric configurations when localized in one of the minima of the double well potential (Fig.~\ref{fig:hsoh}) for a long enough time to measure chirality.

The electric dipole moment and polarizability of HSOH were calculated {\it ab initio} as functions of the $\tau_{\rm HSOH}$ coordinate with the other five semi-rigid internal coordinates fixed at their equilibrium values.
The computed values of the equilibrium dipole moment (in Debye) are $\mu_a=0.053$, $\mu_b=0.744$, and $\mu_c=1.399$, with the $\mu_c$ component having opposite sign for the two enantiomers.
One of the dipole projections ($\mu_a$) has small absolute value, which makes it difficult for the microwave three-wave mixing experiment to do measurements on a time scale faster than the tunnelling.
The computed values of equilibrium polarizability (in atomic units) are $\alpha_{aa}=31.85$, $\alpha_{bb}=26.20$, $\alpha_{cc}=26.51$, $\alpha_{ab}=-0.94$, $\alpha_{ac}=-0.84$, and $\alpha_{bc}=0.07$, where the off-diagonal elements $\alpha_{ac}$ and $\alpha_{bc}$ change sign between two enantiomers.

Using a six-dimensional {\it ab initio} PES~\cite{HSOH_abinit}, we first performed full-dimensional variational calculations of the field-free time independent ro-vibrational energies and wave functions for the rotational quantum number $J=0$ to $J=30$. In order to use these wave functions as basis functions for the time dependent solution, all corresponding ro-vibrational matrix elements of $V(t)$ in Eq.~(\ref{eq1}) and the laboratory fixed electric dipole moment operator $\mu_Z$ were computed.
This was done using the program TROVE~\cite{TROVE,TROVE_curv,TROVE_field}.
The time-dependent wave function exposed by the external field potential $V(t)$ was expressed as a linear combination of the field-free states with the time-dependent coefficients obtained from numerical solution of the time-dependent Schr\"odinger equation using the split operator technique.
The initial wave packets for the two enantiomeric states $\psi_L$ and $\psi_R$ were modelled as a superposition $\psi_{L/R}=1/\sqrt{2}(\phi_{A_1}\pm\phi_{A_2})$ of the two lowest symmetric $\phi_{A_1}$ and antisymmetric $\phi_{A_2}$ torsional eigenstates in the ground rotational state.
We used the laser field $E(t)$ characterized by the carrier wavelength of 800~nm, a Gaussian time profile with FWHM of 0.4~ps and the peak intensity of $1\times 10^8$~V/cm.

\begin{figure}
\includegraphics[scale=0.12]{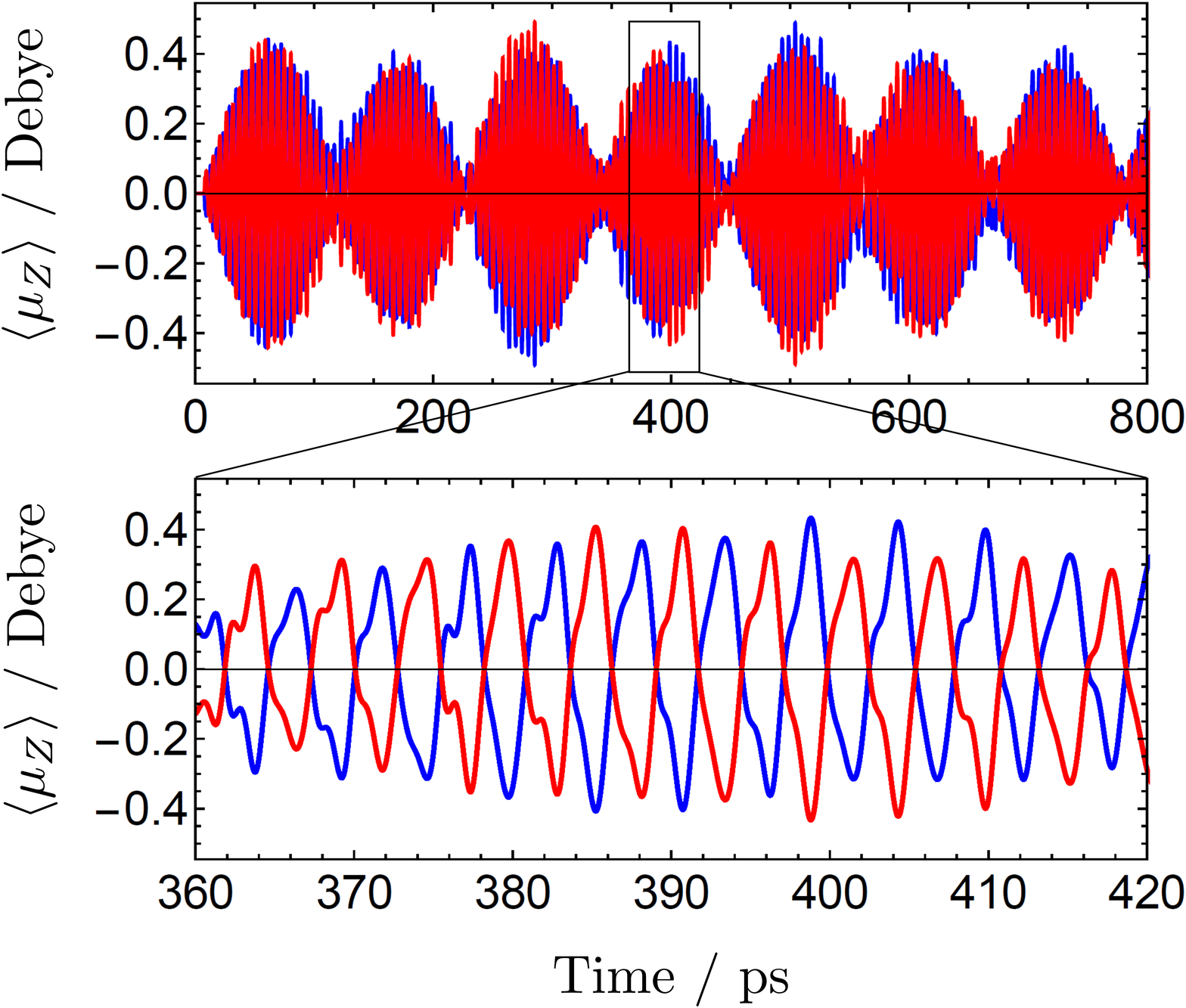}%
\caption{\label{fig:muz}
Calculated time evolution of the electric dipole expectation value ($\mu_Z$) in the laboratory fixed frame, induced by the double-pulse excitation of the enantiomeric states $\psi_L$ and $\psi_R$ (blue and red colors) initially in the ground rotational state.
The time delay between two pulses is set to 6~ps and the pulse mutual polarization angle $\gamma$ is 45$^\circ$.}
\end{figure}

The results of the numerical simulations showing the time evolution of the dipole moment expectation value $\langle\mu_Z\rangle$ are presented in Fig.~\ref{fig:muz}.
As expected, the phases of the dipole oscillations for two enantiomers are shifted by exactly $\pi$ radians.
Such antiphase behaviour will cancel the macroscopic dipole moment of a racemic mixture of two enantiomers, while an excess of one of them will be indicated by non-zero amplitude dipole oscillations.
The corresponding emitted microwave field can then be measured by phase sensitive microwave spectroscopy~\cite{Patterson13,Patterson14}.
From the periodicity of the dipole oscillations in Fig.~\ref{fig:muz}, two major frequencies can be determined, around 0.99~cm$^{-1}$ $\approx B_a+B_b$ and 0.04~cm$^{-1}$ $\approx B_b-B_a$, where $B_a=0.477$~cm$^{-1}$ and $B_b=0.514$~cm$^{-1}$ are rotational constants.

\begin{figure}
\includegraphics[scale=0.16]{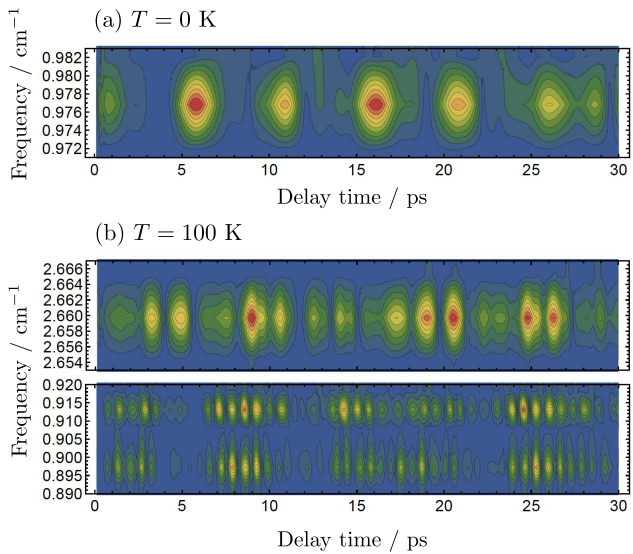}%
\caption{\label{fig:fourier}
Calculated amplitudes of the electric dipole oscillations (as Fourier transform of the dipole time evolution) of the rotational wave packet created by a double-pulse excitation of one of the enantiomers in the ground rotational state ($T=0$~K) and the $T=100$~K thermal equilibrium distribution of ro-vibrational states, plotted against delay time for selected frequency regions.}
\end{figure}

The amplitude of the dipole oscillations can be controlled by tuning the time delay $\tau$ between the excitation pulses.
Fig.~\ref{fig:fourier}.a shows the dipole intensity, obtained as a Fourier transform of the calculated dipole time evolution, plotted against the time delay for the narrow frequency region around $B_a+B_b$.
The intensity exhibits strong maxima at the expected delay times $\tau=5.8$~ps $\approx 1/B$, $10.9$~ps $\approx 2/B$, $16.1$~ps $\approx 3/B$, and $20.8$~ps $\approx 4/B$, etc., where $B=B_a+B_b$. The intensity pattern is not completely symmetric over the revival time, which is due to the interference with other vibrational states included in the simulation.
This property of the dipole oscillations can also be exploited for selective chiral analysis of mixtures by tuning the delay time to match the rotational constants of the molecule of interest.

In order to model a realistic chiral gas at thermal equilibrium, we have simulated 1010 quantum dynamics trajectories corresponding to all initial ro-vibrational states with the Boltzmann thermal factor at $T=100$~K greater than $10^{-6}$. The results of these individual trajectories were then averaged with the corresponding Boltzmann thermal factor. Since our main interest is the temperature effect of the dipole amplitudes, we adopted a simplified model of HSOH as a truly chiral rigid molecule with an impenetrable potential barrier between the two enantiomeric wells.
This approximation does not affect the accuracy of the simulations for low-lying ro-vibrational states, but simplifies the construction of the broken parity enantiomeric states.
Fig~\ref{fig:fourier}.b shows the amplitude of the dipole oscillations for $T=100$~K plotted for various delay times and two selected frequency regions around 0.9~cm$^{-1}$ and 2.6~cm$^{-1}$.
The amplitude shows the oscillatory change against the delay between two pulses, and the peaks are located at different places and are about four times lower compared to the $T=0$~K case (see Fig.~\ref{fig:fourier}.a).
The larger number of peaks and their positions can be explained by the contributions from many initially occupied states, each with different rotational quanta ($J\leq 30$) and thus different amplitude/delay relations.
The density of peaks should grow as the temperature is increased.

\begin{figure}
\includegraphics[scale=0.16]{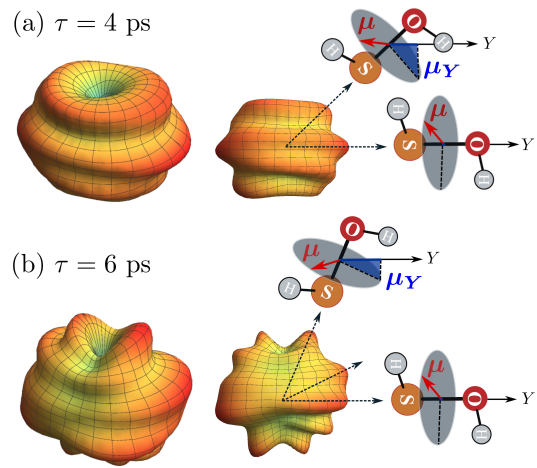}%
\caption{\label{fig:rot}
Probability density functions $P(\theta,\chi)$, averaged over the 0.99~cm$^{-1}$ rotational period, for the delay times (a) $\tau=4$~ps and (b) $\tau=6$~ps. For a simpler visual representation, the densities have been obtained in a simulation with polarization vectors of two excitation pulses set in the $XZ$ plane, which induced an electric dipole polarized along the $Y$ axis.}
\end{figure}

A qualitative picture of the effect of the time delay between the two pulses on the amplitude of the dipole moment can be understood from the reduced angular probability distribution $P(\theta,\chi)=\int dV d\phi\psi^*(t)\psi(t)\sin{\theta}$. Here, $\theta,\chi,\phi$ denote the Euler angles, $dV$ is the volume element associated with the vibrational coordinates, and $\psi(t)$ is the wavepacket. Fig.~\ref{fig:rot} depicts the rotational angular distribution (averaged over the 0.99~cm$^{-1}$ rotational period) after the second pulse for two delay times $\tau=6$~ps and 4~ps, which correspond to large and small dipole amplitudes, respectively. At $\tau=4$~ps (Fig.~\ref{fig:rot}.a), two rotation axes, perpendicular and at about $\angle49^\circ$ to the $Z$ axis, are found to have maximal probabilities with a broad azimuthal distribution.
A rotation about one of these axes causes minor variations of the dipole moment and thus leads to small intensity.
In contrast, a rotational distribution at $\tau=6$~ps (Fig.~\ref{fig:rot}.b) is more localized: it has a distinct maximum at  a small angle to the $Z$ axis of about $25^\circ$ and a more narrow azimuthal distribution.
Hence, a rotation about this axis causes broader oscillations of the dipole and as a consequence stronger intensity.

To summarize, a new scheme for enantiomer differentiation of chiral molecules has been described. The technique employs a pair of linearly polarized intense laser pulses with tilted mutual polarizations and a phase sensitive microwave detector.
A quantum mechanical picture explained the observed anti-phase evolution of the dipole moment for two chiral enantiomers as a consequence of the opposite signs of the off-diagonal anisotropic polarizabilities.
The presented method should be applicable to a wide range of chiral molecules in the ground electronic state with anisotropic polarizability possessing at least one non-zero projection of the dipole moment onto the rotational axes.
Because non-resonant excitation is used it is not necessary to tune the excitation wavelength to the transition of the system, therefore removing the need to cool the sample to temperatures of a few Kelvin. By tuning the microwave detector to the main radiation frequency of the molecule of interest, the method is also suitable for simultaneous chiral analysis of different molecular species and conformers in a mixture.
Moreover, adjusting the delay time between the excitation pulses can selectively amplify the emitted radiation from the molecule of interest.
The ultrafast time scale (several ps) of the presented technique is highly favourable for studying molecules in short-lived chiral states. This is particularly relevant to parity violation experiments~\cite{Fabri15,HiResBook}.

\begin{acknowledgments}

We thank Alec Owens for critically reading the manuscript and suggesting improvements.
This research has been supported by the FP7-MC-IEF project 629237.
We also thank the DiRAC@Darwin HPC cluster, which is the UK HPC facility for particle physics, astrophysics and cosmology and is supported by STFC and BIS.
SY thanks the ERC Advanced Investigator Project 267219 and the COST action MOLIM (CM1405).

\end{acknowledgments}

\end{document}